\documentclass[12pt,twoside]{article}
\usepackage{latexsym,amsfonts}
\renewcommand{\arraystretch}{1.5}

\newcommand{\horisys}[2]{(#1,#2)}
\newcommand{\vertsys}[2]{\left(\begin{array}{c}#1\\#2\end{array}\right)}
\def\bbR{{\mathbb{R}}}

\newcommand{\dfrac}[2]{{\displaystyle{\frac{#1}{#2}}}}
\newcommand{\tfrac}[2]{{\textstyle{\frac{#1}{#2}}}}

\def\LOT{{\rm(lower\ order)}}
\def\image{{\rm image}}

\let\b=\beta

\let\d=\delta
\let\D=\Delta
\let\e=\varepsilon
\let\f=\varphi
\let\F=\Phi

\let\i=\iota

\let\m=\mu

\let\N=\nabla
\let\nd=\nabla

\let\r=\rho
\let\t=\tau
\let\w=\omega

\let\x=\xi

\let\z=\zeta

\newcommand{\de}{\delta}

\newcommand{\om}{\omega}
\renewcommand{\phi}{\varphi}

\newcommand{\si}{\sigma}

\newcommand{\Up}{\Upsilon}

\newcommand{\Rho}{{\mbox{\sf P}}}
\newcommand{\V}{{\mbox{\sf P}}}
\newcommand{\J}{{\mbox{\sf J}}}

\newcommand{\vol}{\mbox{\boldmath $ \epsilon$}}
\newcommand{\bh}{\mbox{\boldmath $ h$}}

\let\s=\sigma
\def\Cal{\mathcal}

\renewcommand{\S}{{S}}
\newcommand{\sS}{{\sf S}}

\newcommand{\lpl}                        
{\mbox{$
\begin{picture}(12.7,8)(-.5,-1)
\put(2,0.2){$+$}
\put(6.2,2.8){\oval(8,8)[l]}
\end{picture}$}}

\def\cA{{\cal A}}
\def\cB{{\cal B}}

\def\cD{{\cal D}}
\def\cE{{\cal E}}
\def\cG{{\cal G}}
\def\ce{\cE}
\def\cF{{\cal F}}
\def\cH{{\cal H}}

\def\cL{{\cal L}}

\def\cP{{\cal P}}
\def\cR{{\cal R}}

\def\cU{{\cal U}}
\def\cV{{\cal V}}
\def\cW{{\cal W}}
\def\cX{{\cal X}}

\def\AD{{\mathbb{D}}}

\def\AX{{\mathbb{X}}}
\def\AY{{\mathbb{Y}}}
\def\AZ{{\mathbb{Z}}}
\def\BD{{\bf D}}
\def\BI{{\bf I}}

\def\bB{{\sf B}}
\def\gG{{\sf G}}
\def\K{\cD_0}

\def\bP{{\mathbb{P}}}

\def\bV{{\mathbb{V}}}
\def\bW{{\mathbb{W}}}
\def\bX{{\mathbb{X}}}
\def\bY{{\mathbb{Y}}}
\def\bZ{{\mathbb{Z}}}

\newcommand{\bg}{{\mbox{\boldmath $g$}}}

\def\ulw{\underline{w}}

\newtheorem{theorem}{Theorem}[section]

\newtheorem{remark}[theorem]{Remark}

\newcommand{\nn}[1]{(\ref{#1})}

\begin{document}

\title{Electromagnetism, metric deformations,
ellipticity and gauge operators on conformal 4-manifolds
}
\markboth{Branson \& Gover}{Gauge operators on conformal 4-manifolds}
\author{Thomas Branson\thanks{Partially supported by NSF grant INT-9724781
and by the Mathematical Sciences Research Institute.}\\
  University of Iowa, USA\and A. Rod Gover\thanks{Partially supported by
the Mathematical Sciences Research Institute.}
\\ University of Auckland, NZ}

\renewcommand{\arraystretch}{1}
\maketitle
\renewcommand{\arraystretch}{1.5}
\begin{abstract}
  On Riemannian signature conformal 4-manifolds we give a conformally
  invariant extension of the Maxwell operator on 1-forms. We show the
  extension is in an appropriate sense injectively elliptic, and
  recovers the invariant gauge operator of Eastwood and Singer. The
  extension has a natural compatibility with the de Rham complex and
  we prove that, given a certain restriction, its conformally
  invariant null space is isomorphic to the first de Rham cohomology.
  General machinery for extending this construction is developed and
  as a second application we describe an elliptic extension of a
  natural operator on perturbations of conformal structure.
  This operator
  is closely linked to a natural sequence of invariant operators that
  we construct explictly. In the conformally flat setting this yields
  a complex known as the conformal deformation complex and for this we
  describe a conformally invariant Hodge theory which parallels the de
  Rham result.
\end{abstract}

\section{Introduction}

Many differential operators that are important in physics and
differential geometry are deficient from the point of view of
ellipticity or hyperbolicity. An example in the setting of Riemannian
4-manifolds is given by a certain natural 4th order conformally
invariant operator on metric perturbations. There are analogues in
higher even dimensions. It seems to us these operators should have an
important role in the relevant deformation theory, and with a view to
applications in this area we were led to consider whether there are
some form of gauge-fixing operators which would extend these to
elliptically coercive
operators. The path to solving this has exposed a rather rich
theory which blends classical elliptic theory with new tools emerging
from representation theory. This enables a systematic approach to a
whole class of problems of this nature. Details of this are developed
in \cite{fup,BrGometric}. Here as a means of introducing and
surveying the key ideas we discuss two examples in 4 dimensions.  By
confining ourselves (for the most part)
to this dimension and these very concrete cases we
are able to present a self-contained treatment using rather elementary
tools. In particular representation theoretic aspects are entirely
suppressed. One of the examples is the above-mentioned operator in
deformation theory and the other is the Maxwell operator of
electromagnetism. Included in the results are conformally invariant,
elliptically coercive
extensions of these operators that lead to a notion of
conformally invariant gauge-fixing. The latter extends and develops the
result in \cite{ES}.
As an application we show that
(given an appropriate restriction) the conformally invariant null space
of the Maxwell extension is precisely the first de Rham cohomology.
Similarly, in the conformally flat case we give a conformally
invariant Hodge theory for the deformation complex.

In the classical theory of electromagnetism, the {\em Maxwell equations}
on a 2-form $F$ over a pseudo-Riemannian 4-manifold are
$dF=0$, $\d F=0$.
Here $\d$ is the formal adjoint of the exterior derivative $d$;
in the original physical problem, the metric has Lorentz signature.
If our manifold is a simply connected region in $\bbR^4$, then
by the Poincare Lemma, the equation $dF=0$ implies that
$F=dA$ for
a 1-form $A$; this is traditionally called the
{\em vector potential} for the Maxwell field $F$.
The Maxwell system then reduces to the single equation
$\delta d A=0$.  Of course the potential is
only determined up to the ``gauge'' freedom of replacing $ A$ with $
A+df$ for some function $ f$.  This ambiguity can be restricted by
imposing further so-called ``gauge-fixing'' equations.  A traditional
choice is the (first order) {\em Lorentz gauge} equation $\delta A=0$. With
this added to the Maxwell equation $\d dA=0$, the vector potential
$ A$ is determined by initial data on a
Cauchy hypersurface.

An important feature of the Maxwell equations is that they are
conformally invariant. This means, among other things, that the
equations are well defined on a {\em conformal space-time}; that is, a
4-manifold equipped with a conformal equivalence class of Lorentzian
metrics, rather than a single
distinguished Lorentzian metric.
The equivalence relation involved is given by
$g\sim \widehat{g}$ iff $\widehat{g}=\Omega^2 g $ for some smooth positive
function $\Omega$.  However, the Lorentz gauge equation is not
conformally invariant, and so is not well defined in the conformal
setting.
(The equation $\d A=0$ is invariant on form-densities of a certain
weight, but not on the form-0-densities where the Maxwell operator
$\d d$ acts.  See below for specifics on densities.)

In \cite{ES}, Eastwood and Singer propose a third-order gauge
fixing operator, which we shall denote $ \sS $,
with principal part $\delta d
\delta$. They show that their operator is not conformally invariant
on general 1-forms, but {\em is} invariant on the conformally invariant
subspace of 1-forms in the kernel of the Maxwell operator $\delta d$.

We shall show that in fact the Eastwood-Singer operator $ \sS$ and the
Maxwell operator $\d d$ can be naturally viewed as parts of a
naturally arising single conformally invariant operator, which we
shall denote by $E$. In fact, in the space-time setting (i.e.\ the
setting of a fixed metric, or {\em scale}), this operator precisely
recovers the system $A\mapsto\horisys{- \delta dA}{{\sS}A}$.
But the important feature of $ E$ is that it has a clear and
well-defined interpretation in the weaker setting of a conformal
structure.  Furthermore, our treatment is directly linked to the
representation theory that gives rise to the natural vector bundles of
conformal geometry.  As a result, it generalises to similar
situations; in particular, that of the so-called {\em metric
  deformation complex} as mentioned above (see also below).  Our treatment,
however, will employ the conformally invariant {\em tractor calculus},
which plays roughly the same role in conformal geometry that
tensor-spinor calculus plays in pseudo-Riemannian geometry.
This calculus, although defined here geometrically, encodes the required
representation theoretic structures and so enables a self-contained treatment
which,  apart from some notational conventions, does not directly appeal to
results from representation theory.

We shall work mainly in the case of Riemannian conformal structures
initially, but all formulas and results on invariant operators
continue in signature to conformal structures of other signatures.
Note that in the discussion above of the Maxwell equations in the
Lorentzian regime, the issue of determination on Cauchy surfaces came
into play.  This is a hyperbolicity property, and to some extent, such
properties tend to correspond to ellipticity properties in the
Riemannian regime.  One concrete way to make the link, and one that
applies in many situations of interest, is to verify that in each
scale there is an operator $T=\horisys{T_1}{T_2}$ and a positive
integer $m$ with the property that
$$
\horisys{T_1}{T_2}\vertsys{{\rm operator}}{{\rm gauge}}
=(\nd^a\nd_a)^m+\LOT.
$$
For example, in the Maxwell-Eastwood-Singer discussion above, we have
$$
\horisys{\d d}{d}\vertsys{\d d}{\d d\d+\LOT}=
(\nd^a\nd_a)^2+\LOT.
$$
This property of being a factor (up to lower-order terms)
of a power the d'Alembertian $\nd^a\nd_a$ in the Lorentzian regime immediately
gives some hyperbolicity properties, while the corresponding property
with the Laplacian $\nd^a\nd_a$ in the Riemannian regime guarantees
elliptic coercivity.

We would like to thank David Calderbank for several  useful conversations.

\section{Elliptically coercive extensions and gauge operators}\label{extend}

In accordance with the remarks directly above, we shall work in the
Riemannian conformal case.  Let $ M $ be a Riemannian 4-manifold with
Levi-Civita connection $\nd $ for the metric $ g$. Let $ \ce^1$ be the
space of smooth sections of the cotangent bundle $ T^*M$. In fact, we
shall often also informally refer to $ \ce^1$ as the cotangent bundle.
Consider the Maxwell equations $d\w=0$, $\d\w=0$ on a 2-form $\w$.
Unless cohomology intervenes, the relation
$d\w=0$ implies that $\w=d\Phi$ for some $1$-form $\F$; as in the
Lorentzian case, we shall call $\F$ the {\em vector potential}.
The
{\em Maxwell operator} $\d d$ on vector potentials fails to be
elliptic -- its leading symbol at a covector $\x$ is $\i(\x)\e(\x)$,
where $\i$ and $\e$ are, respectively, interior and exterior
multiplication -- and this symbol annihilates the range of $\e(\x)$.

Let $\D$ denote the Bochner Laplacian $\nd^a\nd_a$.
For a natural differential operators $P$ on tensors, the condition
that there exists another natural operator (a {\em quasi-inverse})
$Q$ with $QP =\D^m+{\rm(lower\ order)}$ implies that $P$ is
elliptically coercive -- for example, its distributional null space
is finite-dimensional and consists of smooth sections.
If $P$ takes an irreducible bundle to itself,
\cite{ab2} shows that the existence of such a $Q$
is actually equivalent to ellipticity.
One natural notion of ellipticity for operators like the
{\em Euclidean Coulomb gauge}
system $\horisys{\d d}{\d}$ or the Eastwood-Singer system
$\horisys{\d d}{\d d\d+\LOT}$ is the {\em graded ellipticity}
of Douglis and Nirenberg \cite{dn}, which handles block arrays in
which the entries have different orders.  In order not to become
unnecessarily enmeshed in the
technicalities of this, we shall keep as our goal the existence
of a quasi-inverse $Q$ in the sense above.  For purposes of this
paper, we may take the term ``elliptically coercive'', applied
to a natural operator, to mean the existence of a natural quasi-inverse.

Suppose we have a natural differential operator $\cB:\cV_2\to \cV_3$
where $ \cV_i$ is a space of smooth sections of some bundle. As for $
\ce^1$, we shall often refer to such spaces as bundles to simplify the
discussion.  It may be that the realization of $\cB$ in some (and thus
any) conformal scale is elliptic.  If not, we seek a bundle $ \cW$
with $ \cV_3$ as a quotient and an operator $\cX:\cV_2 \to \cW$
such that $ \pi\circ \cX = \cB$, where $ \pi$ is the bundle map $ \pi:
\cW\to \cV_3$. Any (elliptically coercive)
operator $ \cX$ with this property will
be termed an ({\em elliptically coercive}) {\em extension} of $\cB$.

Now suppose there is an operator $\cA:\cV_1\to \cV_2$ such that $
\image(\cA)$ is a subspace of $ \ker(\cB)$. Then we have a sequence
$\cV_1\to \ker(\cB)\subset \cV_2\to \cV_3 $. We are interested in
whether $\ker(\cB)/\image(\cA)$ can be naturally and invariantly
identified with a subspace of $ \ker(\cB)$. Thus we will view $
\image(\cA)$ as the ``gauge freedom'' of the solutions to the $ \cB$
equation.
Let $ \cU$
be the kernel in $ \cW$ of the bundle map $ \pi$. Then we can
write an exact sequence, or equivalently a composition series,
$$
0\to \cU\to\cW\to\cV_3\to 0 \quad \Leftrightarrow \quad \cW= \cV_3 \lpl \cU
$$
to summarise this information about about the filtration of $ \cW$.
Note that when restricted to $ \ker(\cB)$, $ \cX$ takes values in the
subspace $ \cU$. Let us denote the resulting operator $\cG:\ker(\cB)\to
\cU$ and call this the {\em gauge operator} given by  $ \cX$.
The situation so far is summarised in the commutative operator
diagram in Figure \ref{opext}.
Here  $\cP$ is the composition of the gauge
operator $ \cG$ with $ \cA$.
\begin{figure}\label{opext}
%\begin{equation}\label{opext}
%\setlength{\unitlength}{.95pt}
 %       \begin{center}
\begin{picture}(436,200)(-90,-55)

\put(6,0){$\cV_1$}        % V1

\put(36,15){\vector(2,1){110}}     % 1st diagish arrow

\put(34,4){\vector(1,0){15}}       %1st horizontal arrow

\put(140,15){\vector(1,1){15}}     % 3rd diag arrow

\put(56,0){$\ker(\cB)$}        % ker B

\put(137,25){$\cX$}         % H

\put(88,52){$\cP$}           % P

\put(88,25){$\cG$}         % G

\put(86,15){\vector(4,3){60}}      % 2nd diag arrow

\put(95,4){\vector(1,0){15}}       %2nd horizontal arrow

\put(30,-12){$ \cA$}             % A

\put(118,0){$\cV_2$}              % 2nd bun

\put(140,4){$\vector(1,0){15}$}   % 3rd horizontal arrow

\put(144,-12){$\cB$}              % B

\put(170,106){$0$}                 % 0

\put(173,100){\vector(0,-1){10}}  % 0 to U down vector

\put(170,68){$\cU$}               %U

\put(173,62){\vector(0,-1){10}}  % U to W down vector

\put(170,34){$\cW$}              % W

\put(173,28){\vector(0,-1){10}}  % W  to V3 down vector

\put(170,0){$\cV_3$}

\put(173,-6){\vector(0,-1){10}}

\put(170,-34){$0$}

\end{picture}
%\end{center}
%\setlength{\unitlength}{1pt}
%\end{equation}
\caption{The extension $ \cX$ of $ \cB$ is a gauge extension
if $ \cP$ is elliptic.}
\end{figure}

In the best of worlds $ \cX$ could turn out to be what might be called
a ``gauge-fixing extension'' of $ \cB$ (relative to $ \cA$). This would
mean for any $ v_2 \in \ker(\cB)$ there is $ v_1\in \cV_1$ such that $
\cX(v_2+\cA(v_1))=0$. That is, there should exist $ v_1 $ solving $\cP
(v_1)= -\cG(v_2) $. The remaining freedom in $ v_2+\cA(v_1)$ is then
reduced to adding some $u_1 $ from $\ker(\cP)$. It is ideal if this
remaining freedom has no impact on the quotient
$\ker(\cB)/\image(\cA)$. That is, a gauge-fixing extension should have
$ \ker(\cP)\subset \ker(\cA)$.
%Of course the gauge extension of $ \cB$ is
%dependent on $ \cA$ so we should really describe this a gauge
%extension of $\cB$ relative to $ \cA$. Since $ \cA$ is clear in the
%cases we consider we will avoid this lengthy expression.
Whether a particular extension is gauge-fixing in this way is in
general a non-local problem.  One class of extensions in which there is
at least a good chance arises when $ \cP$ is elliptic. Let us term such an
extension $ \cX$ a {\em gauge extension} of $ \cB$ if $ \cP:\cV_1\to
\cU$ is elliptic. We will say the kernel of $\cP$
is {\em harmless} if $  \ker(\cP)\subset \ker(\cA)$.

It is useful to note what all of these objects are in the
Maxwell setting.
Here $\cB $ is the Maxwell
operator $ \d d: \ce^1\to \ce^1$ and $ \cA$ the exterior derivative $
d:\ce\to \ce^1$, where we write $ \ce$ for (the smooth sections of)
the trivial bundle. As in the space-time case, the traditional choice
for $\cG$ has been the divergence $\d$; in Riemannian signature this
is called the (Euclidean) {\em Coulomb gauge}.
%\edz{Tom: Lorentz
%gauge versus Coulomb gauge as terminology. Any comments?  {\sf I think this
%is about right, as you have it.  In the Lorentz setting, the ``Coulomb
%gauge'' refers to $\d A=0$ and $A_0=0$ (relative to some space-time
%splitting).}}
So then $
\cX$ is just the operator $ (\d d, \d):\ce^1 \to\ce^1\oplus \ce $.
Observe that this is a gauge extension of the Maxwell operator, as
the $\cP$ of Figure \ref{opext} can be taken to be $-\Delta$.
(In this connection, note that $-\Delta=\d d$ on functions.)
%
%\edz{In final comments of paper maybe say why we
% cannot fix this to be conformally invariant}
%
To see that this is gauge fixing in the sense we have described, we
need to know the gauge can be attained. In the compact setting, for
example, this is always possible because to reach the Coulomb gauge
involves solving $\d(\F+df)=0$ for $f$; that is, solve $\D f=\d \F$.
 This
achievable by standard elliptic theory.  In addition, the operator $
(\d d, \d)$ is an elliptically coercive extension, since
composing with $(1,d): \ce^1\oplus \ce\to \ce^1 $ yields $\d d+d\d$,
the form Laplacian (which agrees with $-\D$ up to lower order terms).
The problem here is just that the operator $(\d d, \d)$ does not
correspond (in the sense of the discussion of Figure \ref{opext}) to
any conformally invariant operator.

On a Riemannian 4-manifold the Eastwood-Singer gauge
operator may also be viewed as giving an extension of the form $(-\d d, \sS):
\ce^1 \to \ce^1 \oplus \ce$.
%Recall the gauge operator $ \sS$ has principal part
%$\d d \d $ so the ellipticity can be verified by composing with $(\d
%d, d) $ which yields $\D^2+\LOT$.  Attaining the ES gauge involves
%solving an equation of the form $\D^2 f = - \d U \F $ which again is
%always possible.  \edz{Rod: Check notation etc here:
%R Qez: Same ques as above. Really true? What
%  restrictions? I know compact is fine . . .}
In the conformal setting, however, the replacement for $\ce^1 \oplus \ce
$ does not split as a direct sum.
It turns out that there is a conformally
invariant elliptically coercive
gauge extension of Maxwell operator, $ E:\ce^1 \to
\ce_{\bar{A}}[-3]$ where $ \ce_{\bar{A}}[-3]$ (cf.\ $ \cW$ in
Figure \ref{opext})
is a bundle with the composition series $\cF=\ce^1[-2] \lpl
\ce[-4] $ (cf.\ $ \cW=\cV_3\lpl \cU$), $ \ce^1[-2]$ is a bundle
isomorphic to the cotangent bundle, and $ \ce[-4]$ is
isomorphic to the trivial bundle. In a choice of metric from the
conformal class, $ E$ recovers the operator $(-\d d, \sS)$.

\section{Conformal geometry and tractor calculus} \label{conf}

%Note that basically everything that has been said so far has an
%analogue in arbitrary dimensions $n$ which is just as easy to state.
%In particular, the system $d\w=0$, $\d\w=0$ on an $\ell$-form in
%dimension $n=2\ell$ enjoys the same conformal invariance property as
%its 4-dimensional special case, and there is a generalisation of the
%Eastwood-Singer gauge to this setting \cite{fup}. \edz{MOVE ME: Tom: where do
%  you want to stick this remark. I don't think it belongs here. I have
%  alluded to our work a bit in the intro but maybe this somehow
%  belongs in there or else in the Maxwell section.}
Tractor calculus is a conformally invariant calculus based on natural
bundles in conformal geometry. It includes the Cartan connection
manifested as an induced vector bundle connection.  However, one of the
main misunderstandings in the area is that this is the end of the
story.  There are several other fundamental and equally important
invariant operators involved. Perhaps even more importantly, the
calculus provides the right forum for using results and ideas from
representation theory in the ``curved''
(for example, conformally
curved)
%\edz{{\sf Is this what we mean?  Alternatively, we could say
%``especially conformally curved''.}}
differential geometric setting.
Although this aspect has been suppressed in the current article, we
should point out that it has been very influential in this work and
will be described in \cite{fup,BrGometric}.  We summarise here some
key tools of tractor calculus. This is mainly drawn from the development
presented in \cite{Cap-Gover2}, but many of the ideas and tools had
their origins in \cite{T}, \cite{BEGo}, and \cite{Goadv}.  The
notation and conventions in general follow the last two sources.

For the remainder of this section
there is no real advantage in restricting to dimension 4, so
we work on a real conformal $n$-manifold $M$, where $n\geq 3$.
%\edz{{\sf Have to be careful here, as $\d d$ is not invariant on
%1-forms when we leave dim.\ 4.  The analogue is $(\ell-1)$-forms, where
%$\ell=n/2$.  I'll alter things accordingly just below; shouldn't
%be too bad.  We could consider this the moved remark from above --
%i.e.\ there are analogous objects in other dimensions.}
%Okay -- I'll watch that when I proof-read.
%There shouldn't be too much Maxwell stuff in this section}
That is, we have a pair $(M,[g])$, where $M$ is a smooth $n$-manifold
and $[g]$ is a conformal equivalence class of Riemannian metrics. (In
fact {\em most} results in this section are signature independent.)
Two metrics $g$ and $\widehat{g}$ are said to be {\em conformally
  equivalent} if $\widehat{g}=\Omega^2 g$ for some positive smooth
function $\Omega$. (The replacement of the metric $ g$ by the metric $
\widehat{g}$ is called a {\em conformal transformation}.)  For a given
conformal manifold $(M,[g])$, we shall denote by $\Cal Q$ the bundle
of metrics.  $\Cal Q$ is a subbundle of $S^2T^*M$ with fibre $\bbR_+$.
From this principal bundle
%
%\edz{{\sf 10/23 Seems to be something added here with} Bbb; {\sf I changed
%to} mathbb {\sf to avoid that error msg.  More importantly, there now
%seems to be redundancy -- we are saying twice what the fibre is.}}
%
there are natural line bundles $ \ce[w]$, $w\in\bbR$, on $ (M,[g])$
induced from the irreducible representations of $\bbR_+$.  A section
of $\ce[w]$ corresponds to a real-valued function $ f$ on $\Cal Q$
with the homogeneity property $f(\Omega^2 g, x)=\Omega^w f(g,x)$.
%, where
%$\Omega$ is a positive function on $M$, $x\in M$, and $g$ is a metric
%from the conformal class $[g]$.
%\edz{{\sf Commented out the wording where we rolled out the
%boilerplate on $\Omega$
%yet again.}}

For many discussions it will be convenient to use Penrose's abstract
index notation. Thus for example we will sometimes use $\ce_a$ as an
alternative notation for the cotangent bundle
$ \ce^1$ or its smooth sections.
%(which recall means the cotangent
%bundle or its smooth sections according to context).
%\edz{{\sf Again commented some repeated language.}}
We then write $\ce_{ab}$ for $ \otimes^2 \ce_a$, $ \ce_{(ab)}$ for the
symmetrisation of this and so forth.  Similarly $ \ce^a$ indicates the
tangent bundle or its smooth sections. An index which appears twice,
once raised and once lowered, indicates a contraction.  These
conventions will be extended in an obvious way to the tractor bundles
described below. In all settings, indices may also be omitted if the
meaning is clear.  We use the notation $\ce_a[w]$ for $\ce_a\otimes
\ce[w]$ and so on.

With $\ce_+[-2]$ denoting the fibre subbundle of $\ce[-2]$
corresponding to $\bbR_+\subset\bbR$, it is easily verifed
that $\ce_+[-2]$ is canonically isomorphic to $\Cal Q$. The {\em
  conformal metric\/} ${\mbox{\boldmath $g$}}_{ab}$ is the
tautological section of $\ce_{ab}[2]$ that represents the map
$\ce_+[-2]\cong{\Cal Q}\to\ce_{(ab)}$. Then $\bg^{ab}$ is the section
of $\ce^{ab}[-2]$ such that $\bg_{ab}\bg^{bc}=\de_a{}^c$, the identity
endomorphism on $ \ce_c$.  The conformal metric and its inverse
 will be used to raise and lower indices without further
mention.  Given a choice of metric $ g$ from the conformal class, we
write $ \nabla_a$ for the corresponding Levi-Civita connection. With
these conventions the Laplacian $ \Delta$ is given
$\Delta=\bg^{ab}\nd_a\nd_b= \nd^b\nd_b\,$. In view of the isomorphism
$\ce_+[-2]\cong{\Cal Q}$, a choice of metric also trivialises the
bundles $ \ce[w]$.
%In particular we will write $\xi^g$ for the
%canonical section of $ \ce[1]$ satisfying $g=(\xi^g)^{-1}\bg $.
%Conversely a choice of non-vanishing section of $ \ce[1]$ clearly
%determines a metric by this relation, so such a $ \xi^g$ is termed a
%choice of conformal scale.
This determines a connection on $ \ce[w]$
via  the exterior
derivative on functions. We shall also denote such a connection by $
\nabla_a$ and refer to it the Levi-Civita connection.
%Note in
%particular then that, by definition, $ \nd_a \xi^g=0$, so $ \nd_a$
%also preserves the conformal metric.
Defined in
this way  the Levi-Civita connection preserves the conformal metric.

The Riemannian curvature is defined by
$
(\nd_a\nd_b-\nd_b\nd_a)v^c=R_{ab}{}^c{}_dv^d ,
$
on tangent vector fields $ v$.
This can be decomposed into the totally trace-free Weyl curvature
$C_{abcd}$ and a remaining part described by the symmetric {\em
  Rho-tensor} $\Rho_{ab}$, according to
$$
R_{abcd}=C_{abcd}+2\bg_{c[a}\Rho_{b]d}+2\bg_{d[b}\Rho_{a]c},
$$
where $[\cdots]$ indicates the antisymmetrisation over the enclosed indices.
The Rho-tensor is a trace modification of the Ricci tensor $R_{ab}$. We write
$ \J$ for the trace $ \V^a{}_{a}$ of $ \V$.

 Under a conformal transformation the Levi-Civita connection
  then transforms as follows:
\begin{equation}\label{transform}
\widehat{\nd_a u_b}=\nd_a u_b -\Up_a u_b-\Up_b u_a +\bg_{ab} \Up^c u_c, \qquad
 \widehat{\nd_a \si} = \nd_a \si +w\Up_a \si.
\end{equation}
Here $ u_b\in \ce_b$, $ \si\in \ce[w]$, and
$\Upsilon_a:=\Omega^{-1}\nabla_a\Omega$.

Specialising for the moment to dimension 4, on a 2-form $\om $,
Maxwell's equations $d\w=0$, $\d\w=0$ may alternatively be written as
$3\nd_{[a} \w_{bc]}=0$ and $ -\nd^c\w_{ca}=0$ respectively.
Similarly the {\em Maxwell operator} on vector potentials is the
operator $\d d: \F_a\mapsto -2\nd^b(\nd_b \F_a-\nd_a \F_b)$. It is
straightforward to verify directly, using the transformation formulae,
that these equations are conformally invariant. On the other hand for
$\phi\in\ce_a[w] $ we have
$$
\widehat{\nd}^a\phi_a =\nd^a\phi_a+(w+2)\Up^a\phi_a .
$$
%
%\edz{Tom: check changes here. I needed to put some weights in for your
%  argument that ``$\d$ is invariant
%$\cE_a[-2]\to\cE[-4]$'' {\sf OK}}
%
This shows that the Coulomb gauge operator $\d$ is invariant
$\cE_a[-2]\to\cE[-4]$ but {\em is not} conformally invariant on
$\cE_a$.  In fact it is clearly not even invariant on the subspace
of exact 1-forms.
Thus this is incompatible with the invariance of the Maxwell
operator $\d d$, which acts invariantly on $\cE_a=\cE_a[0]$.

A natural generalisation of the Maxwell equations to
even dimensions $n=2\ell$ is
the system $d\om=0$, $\d\om=0$ on $\cE_{a_1\cdots a_\ell}\,$.
Again, unless cohomology intervenes, there is a vector potential
$\F\in\cE_{a_1\cdots a_{\ell-1}}\,$, and the equations reduce to
the invariant equation $\d d\F=0$.  Appending the Coulomb gauge
equation $\d A=0$ in a scale, we have an elliptically coercive system,
but again this system is not conformally invariant:
$\d$ carries $\cE_{a_1\cdots a_{\ell-1}}[-2]$
to $\cE_{a_1\cdots a_{\ell-2}}[-4]$ invariantly, but does not act
invariantly on $\cE_{a_1\cdots a_{\ell-1}}[0]$.  There is, however,
an analogue
of the invariant Eastwood-Singer gauge \cite{fup}.
%\edz{{\sf Here was our remark on other dimensions, and also keeping
%our promise that the remainder of this section would be in all
%dims.}}

The Weyl curvature is
conformally invariant, that is
%$\widehat{C}_{abcd}=C_{abcd}$,
$\widehat{C}_{ab}{}^c{}_d=C_{ab}{}^c{}_d$,
and the
Rho-tensor transforms by
\begin{equation}\label{Rhotrans}
\textstyle \widehat{\V}_{ab}=\V_{ab}-\nd_a \Up_b +\Up_a\Up_b
-\frac{1}{2} \Up^c\Up_c \bg_{ab} .
\end{equation}

%For the the density bundle $\ce[1]$, we have the jet exact sequence at
%2-jets,
%$$
%0\to \ce_{(ab)}[1]\to J^2(\ce[1])\to J^1(\ce[1])\to 0,
%$$
%where $ (\cdots)$ indicates symmetrisation over the enclosed
%indices.  Note we have a bundle homomorphism $ \ce_{(ab)}[1] \to \ce[-1]$
%given by complete contraction with $\bg^{ab}$. This is split via
%$ \rho\mapsto  \frac{1}{n}\rho\bg_{ab}$ and so
%the conformal structure decomposes $\ce_{(ab)}$ into the
%direct sum $\ce_{(ab)_0}[1]\oplus\ce[-1]$.
Let us write $\ce_{(ab)_0}[1]$ for the symmetric trace-free part of $
\ce_{ab}[1]$. Then $\ce_{(ab)_0}[1]$ is naturally a smooth subbundle
of the bundle of 2-jets $J^2(\ce[1])$ of the density bundle $\ce[1]$.
The {\em
  standard tractor bundle} $\ce^A$ is defined by the exact sequence
\begin{equation}\label{trdef}
0\to \ce_{(ab)_0}[1]\to J^2(\ce[1])\to \ce^A\to 0.
\end{equation}
{}The jet exact sequence at 2-jets and the corresponding sequence at
1-jets, viz.\ $ 0\to \ce_{a}[1]\to J^1(\ce[1])\to \ce[1]\to 0 , $
determine a composition series for $\ce^A$ which we can summarise
via the semi-direct sum notation by $ \ce^A= \ce[1]\lpl
\ce_a[1]\lpl\ce[-1]$. We denote by $ X^A$ the canonical section of $
\ce^A[1]:=\ce^A\otimes \ce[1]$ corresponding to the mapping $ \ce[-1]
\to \ce^A$.
%Composing the canonical projection $J^2(\ce[1])\to\ce^A$ with the
%2-jet operator $ j^2$ yields an invariant differential operator
%$\tfrac1{n}D^A:\ce[1]\to \ce^A$.
%On the other hand, if we choose a
%metric $ g$ from the conformal class, then the map
%$$
%j^2_x\si \mapsto [\tfrac1{n}{D^A\si}(x)]_g:=(\si(x),\nabla_a\si(x),
%-\tfrac{1}{n}\bg^{ab}(\Delta+\J)\si(x))
%$$
A choice of metric from the equivalence class determines an isomorphism
$\ce^A\to
\ce[1]\oplus\ce_a[1]\oplus\ce[-1]=:[\ce^A]_g$ of vector bundles.
If the image
of $ V^A\in\ce^A$ is $[V^A]_g=(\si,\mu_a,\tau) $, then for $\hat{g}=
\Omega^2 g$ we have
$$
[V^A]_{\hat{g}}=\widehat{(\si,\mu_a,\tau)}=(\si,\mu_a+\si\Up_a,\tau-
\Up_b\mu^b- \tfrac{1}{2}\si\Up_b\Up^b).
$$
This transformation formula characterises sections of $\ce^A$ in terms
of triples in $ \ce[1]\oplus\ce_a[1]\oplus\ce[-1]$ at all possible
scales.
In this
notation $ [X^A]_g=(0,0,1)$. It is convenient to introduce
scale-dependent sections $Z^A{}^b\in\ce^{Ab}[-1]$ and
$Y^A\in\ce^A[-1]$ mapping into the other slots of these triples so
that $[V^A]_g=(\si,\mu_a,\tau)$ is equivalent to
\begin{equation}\label{Horiz}
V^A=Y^A\s+Z^{Ab}\m_b+X^A\t.
\end{equation}

The standard tractor bundle has an invariant metric $ h_{AB}$ of
signature $(p+1,q+1)$ and an invariant connection, which we shall also
denote by $ \nabla_a$, preserving $h_{AB}\,$. If $ V^A$ is as above
and $\underline{V}^B \in \ce^B$ is given by
$[\underline{V}^B]_g=(\underline{\s},\underline{\m}_b,
\underline{\t})$, then
$$
h_{AB}V^A\underline{V}^B=\m^b\underline{\m}_b+\s\underline{\t}+
\t\underline{\s}.
$$
Using $h_{AB}$ and its inverse to raise and lower indices, we
immediately see that
$$
Y_AX^A=1,\ \ Z_{Ab}Z^A{}_c=\bg_{bc},
$$
and that all other quadratic combinations that contract the tractor
index vanish.
In fact the metric may be decomposed into a sum of projections, $
h_{AB}=Z_A{}^cZ_{Bc}+X_AY_B+Y_AX_B\,$.
The tractor metric will be used to raise and lower indices without
further comment.  We shall use either ``horizontal'' (as in
$[V^A]_g=(\si,\mu_a,\tau)$ or \nn{Horiz})
or ``vertical'' (as in \nn{stdtracconn} below)
notation, depending on which is clearer in each given situation.

If, for a metric $ g$ from the conformal class, $V^A \in\ce^A$ is given by
$[V^A]_g=(\si,\mu_a,\tau)$, then the invariant tractor  connection is given by
\renewcommand{\arraystretch}{1}
\begin{equation}\label{stdtracconn}
[\nabla_a V^A]_g =
\left(\begin{array}{c} \nabla_a \si-\mu_a \\
                       \nabla_a \mu_b+ \bg_{ab} \tau +\V_{ab}\si \\
                       \nabla_a \tau - \V_{ab}\mu^b  \end{array}\right) .
\end{equation}
\renewcommand{\arraystretch}{1.5}

Tensor products of the standard tractor bundle, skew or symmetric
parts of these, and so forth are all termed tractor bundles.
The
bundle tensor product of such a bundle with $ \ce[w]$, for some real
number weight $ w$, is termed a weighted tractor bundle.  Given a  choice
of conformal  scale we have the corresponding Levi-Civita connection
on tensor and density bundles.  In this
setting we can use the coupled Levi-Civita tractor connection to act
on sections of the tensor product of a tensor bundle with a tractor
bundle. This is defined by the Leibniz rule in the usual way.
In particular  we have
\begin{equation}\label{connids}
\begin{array}{rcl}
\nd_aX_A=Z_{Aa}\,, &
\nd_aZ_{Ab}=-\V_{ab}X_A-Y_A\bg_{ab}\,, & \nd_aY_A=\V_{ab}Z_A{}^b ,
\end{array}
\end{equation}
which are useful for calculations.

The adjoint tractor bundle $
 \ce^\alpha$ is simply the second exterior power of the tractor
 bundle, i.e.  $\ce^\alpha:=\ce^{[AB]}$.  It follows that it has a
 composition series
$$
\ce^a\lpl(\ce\oplus\ce_{[ab]}[2])\lpl\ce_a\,.
$$
Given a choice of metric, this decomposes so that the semi-direct sum becomes
a direct sum (i.e. $\lpl$ gets replaced by $\oplus $),
and it is convenient to write
sections $ \bV^\beta $ of $ \ce^\beta$ as corresponding 4-tuples
$$
[\bV^\alpha]_g=
(\xi^a,\Phi_b{}^a,\phi,\omega_a).
$$
Under a conformal transformation $ g\mapsto
\hat{g}$, we have
$$
\begin{array}{l}
[\bV^\alpha]_{\hat{g}}=
\widehat{(\xi^a,\Phi_b{}^a,\f,\omega_a)}= \\
(\xi^a,\Phi_b{}^a+\xi^a\Upsilon_b-
\xi_b \Upsilon^a,\f+\xi^a\Up_a,\omega_a-\Phi_a{}^b
\Upsilon_b-\f\Up_a
-\xi^b\Upsilon_b\Upsilon_a+
\tfrac{1}{2}\xi_a\Upsilon_k\Upsilon^k).
\end{array}
$$

We can view the adjoint tractor bundle as the bundle of filtration and
metric-preserving endomorphisms of the standard tractor bundle, and we
take one-half of the trace form as the inner product $B_{\alpha\beta}$
on $ \ce^\beta$. (The typical fibre of $ \ce^\alpha$  is the Lie algebra
${\mathfrak{so}}(n+1,1)$.)  That is if $[\bV]_g=
(\xi^a,\Phi_b{}^a,\f,\omega_a)$ and $[\underline{\bV}]_g=
(\underline{\xi}^a,\underline{\Phi}_b{}^a,\underline{\f},
\underline{\omega}_a)$, then
$$
B_{\alpha\beta}\bV^{\alpha}
\underline{\bV}^\beta=\frac12\F_a{}^b\underline{\F}_b{}^a
+\f\underline{\f}+\xi^a\underline{\r}_a+\r_a\underline{\xi}^a.
$$

The connection on the standard tractor bundle gives a connection on its
tensor powers by the Leibniz
rule, and in particular on $ \ce^\beta$.
For a section $\bV^\alpha$ of $\ce^\alpha$
with $ [\bV^\beta]_g=(\xi^b,\Phi_c{}^b,\f, \omega_b)$, this  is given by
\renewcommand{\arraystretch}{1}
\begin{equation}\label{adjtracconn}
[\nabla_a\bV^\beta]_g = \left(\begin{array}{c} \nabla_a \xi^b-\Phi_a{}^b
-\d_a{}^b\f \\
                       \nabla_a \Phi_c{}^b+ \delta_a{}^b \omega_c
-\bg_{ac}\omega^b+\xi^b\V_{ac}-\xi_c\V_a{}^b \\
\nd_a\f+\w_a+\xi^k\V_{ka} \\
\nabla_a \omega_b-\V_{ka}\Phi_b{}^k-\V_{ba}\f  \end{array}\right)
\end{equation}
\renewcommand{\arraystretch}{1.5}

Alternatively, in analogy with the standard tractor calculations above,
we can write
$$
\bV^\b=\bY^\b{}_a\xi^a+\bZ^\b{}_a{}^b\F_b{}^a+\bW^\b\f+\bX^{\b a}\w_a\,,
$$
where $\bX^{\b a}$ is an invariant section, and $\bY^\b{}_a\,$,
$\bZ^\b{}_a{}^b$, and $\bW^\b$ are scale-dependent sections.  It is
straightforward to write formulae for $ \nd$ on these (cf.\
\nn{connids}).

We conclude with some observations we will need later. One is that the
Yamabe operator extends to a conformally operator on tractor bundles
of the appropriate weight. That is there is a conformally invariant
differential operator $ \Box: \ce^\Psi[1-n/2]\to \ce^\Psi[-1-n/2]$, where
$ \ce^\Psi[w]$ indicates any tractor bundle of weight $ w$. This is given by
the usual formula,
\begin{equation} \label{box}
\Box V :=\nd_p\nd^p V+w\J V,
\end{equation}
except now $ \nd$ indicates the coupled tractor-Levi-Civita connection.

Now consider $
\bbR^{n+2}$ equipped with an inner product $ \bh$ of signature $
(n+1,1)$.  The space of null lines is a quadric in the
projectivisation $\bP_{n+1}=\bP\bbR^{n+2}$ with a
(conformally flat) conformal structure. This $n$-sphere is usually
regarded as the standard flat model for a conformal structure and we
will refer to this as the conformal sphere. The orthogonal group $
G:={\rm O}(\bh)$ acts conformally on this space which may be
identified with $ G/P$, where $ P$ is a certain parabolic subgroup of $
G$.  Now $ G$ is a principal $ P$-bundle over $ G/P$ and in this
setting the standard tractor bundle is induced from the defining
representation of $ G$ regarded as a $ P$-module.  Since this space
carries a representation of $ G$, the bundle is trivialisable.  It
follows easily from the normality of the tractor connection
(see \cite{Cap-Gover2} and \cite{Cap-Gover})
 that under this trivialisation the
operator $ \Box$ agrees with the trivially extended Yamabe operator.
Thus $ \Box$ is elliptic in this flat model but
therefore also in general.

\section{Gauge extension of the Maxwell operator}

The exterior derivative operators are well defined
diffeomorphism-invariant operators on any smooth manifold, and so in
particular are well defined on a conformal manifold. However there are
other conformally invariant operators between forms.  On the dimension
4 conformal sphere
(section \ref{conf}) the following diagram gives all $G$-invariant
operators \cite{ERice} between the forms (via the isomorphisms $
\ce^1[-2]\cong \ce^3$ and $ \ce[-4]\cong \ce^4$; see immediately below).
In fact these are
the only $ G$-invariant operators on forms which take values in
irreducible tensor bundles.
\begin{equation}\label{maxbgg}
\setlength{\unitlength}{.95pt}
\begin{picture}(336,120)(58,-55)
\put(49,-20){\line(0,-1){30}} 
\put(49,-50){\line(1,0){295}}
\put(344,-50){\vector(0,1){30}}
\put(43,0){$\ce$}  
\put(90,0){$\ce^1$} 
\put(100,16){\line(0,1){36}}  
\put(100,52){\line(1,0){179}}
\put(279,52){\vector(0,-1){37}}
\put(184,28){$\ce_{+}^{2}$} 
\put(184,-32){$\ce_{-}^{2}$}
\put(255,0){$\ce^1[-2]$}    
\put(330,0){$\ce[-4]$}    
\put(63,4){\vector(1,0){15}} 
\put(140,15){\vector(1,1){15}}   
\put(140,-15){\vector(1,-1){15}} 
\put(235,29){\vector(1,-1){15}}  
\put(235,-29){\vector(1,1){15}}  
\put(300,4){\vector(1,0){15}}    
\end{picture}
\setlength{\unitlength}{1pt}
\end{equation}
Here $\ce_{\pm}^{2}$ are the self-dual and anti-self-dual 2-forms.
Proceeding from the left, the first short horizontal operator is the exterior
derivative on functions.  The first diagonal operators
are given by the exterior derivative followed with projections into
$\ce^{2}_{\pm}$ and the remaining short arrow operators are formal adjoints
of these. (Formal adjoints are with respect to the conformally
invariant inner product of section \ref{hodge} below.)
 The operator $ \ce^1\to \ce^1[-2]$ is of course the Maxwell
operator $ \d d$ (which is up to scale is the composition around either
edge of the diamond) and the longest operator has principal part $
\Delta^2$. The generalisation of these to invariant operators on
general conformal 4-manifolds is straightforward except for the last of these,
which in that generality is known as the Paneitz operator.
This operator, which we shall denote $P_4$, is given by the formula
$$
P_4:=\nd_b(\nd^b\nd^c+4\V^{bc}-2\J\bg^{bc})\nd_c :\ce \to \ce[-4].
$$
The Paneitz operator is formally self-adjoint and annihilates
constant functions.
Among  operators with these
properties it is known to be the unique (up to constant multiples)
conformally
invariant natural operator between these bundles.
Most of the operators from diagram \nn{maxbgg} play a role in the gauge
extension of the Maxwell operator.

Let us temporarily work in a general dimension $n\geq 3$.
Recall that the standard tractor bundle has a composition series
$$
\ce_A=\ce[1]\lpl \ce_a[1]\lpl\ce[-1]
$$
and the bundle injection $ \ce[-1]\to \ce_A$ is given by $
\rho\mapsto \rho X_A$. Let us denote by $ \ce_{\bar{A}}$ the quotient
of $ \ce_A$ by the image of this map, and
let $\ce^{\bar{A}}$ be the dual bundle.
Extending the conventions from
above, we write $ \ce_{\bar{A}}[w]$ to mean $ \ce_{\bar{A}}\otimes
\ce[w]$ and so forth.
Clearly $\ce_{\bar{A}}[w-1]$ has the composition series $ \ce_{\bar{A}}[w-1]=
\ce[w]\lpl\ce_a[w]$.
We define
$$
\begin{array}{rl}
I_{\bar{A}}{}^a:\ce_a[w]&\to
\ce_{\bar{A}}[w-1], \\
\m_a&\mapsto I_{\bar{A}}{}^a\m_a
\end{array}
$$
to be the canonical inclusion.  Given choice of metric $g$, we have $
[\ce_{\bar{A}}[w-1]]_g= \ce[w]\oplus \ce_a[w]$ and the inclusion is given by
$ [I_{\bar{A}}{}^a\mu_a]_g=(0,\mu_a) $.

Now provided $ w\notin\{ 1-n/2,2-n\}$, the algebraic bundle surjection
$P_{\bar{A}}{}^A: \ce_A[w-1]\to \ce_{\bar{A}}[w-1]$ has an invariant
differential splitting. That is, there is an operator
\begin{equation}\label{diffsplit1}
\S_A{}^{\bar{A}}:\ce_{\bar{A}}[w-1]\to\ce_A[w-1]
\end{equation}
 such that the
composition $ P_{\bar{B}}{}^A\S_A{}^{\bar{A}}$ is the identity $
\delta_{\bar{B}}{}^{\bar{A}}$ on $ \ce_{\bar{A}}[w-1]$.
In terms of the
decomposition $[\ce_A[w-1]]_g=\ce[w]\oplus \ce_a[w] \oplus \ce[w-2] $
this  is given by
$$
\left(\begin{array}{c} \si \\
                       \mu_a \end{array}\right)
 \mapsto
 \left(\begin{array}{c} \si \\
                        \mu_a \\
-\dfrac{1}{n+2w-2}\left(\dfrac1{n+w-2}\D+\J\right)\si-\dfrac{1}{n+w-2}\nd^b\m_b
\end{array}\right).
$$
In the alternative notation,
$$
\z_{\bar{A}}=Y_{\bar{A}}\s+Z_{\bar{A}}{}^a\m_a
$$
is carried to
$$
\begin{array}{l}
\S_A{}^{\bar{A}}\z_{\bar{A}}=Y_AX^{\bar{A}}\z_{\bar{A}}
+Z_A{}^aZ^{\bar{A}}{}_a\z_{\bar{A}} \\
{}-\dfrac{1}{n+2w-2}X_A\left(\dfrac1{n+w-2}\D+\J\right)(X^{\bar{A}}
\z_{\bar{A}})-\dfrac{1}{n+w-2}X_A\nd^b(Z^{\bar{A}}{}_b\z_{\bar{A}}).
\end{array}
$$
Here $Y_{\bar{A}}$ is the image of $ Y_A$ under
$\ce_{A}[-1]\to\ce_{\bar{A}}[-1]$, $ X^{\bar{A}}$ the section of $
\ce^{\bar{A}}[1]$ with image $X^A$ under $\ce^{\bar{A}}[1]\to \ce^A[1]
$ and so forth.

We would now like to introduce the formal adjoints
$\bar{\S}^{\bar{A}A}:\ce_A[-3]\to \ce^{\bar{A}}[-3]$ and
$\bar{I}^a{}_{\bar{A}}:\ce^{\bar{A}}[-3]\to \ce^a[-4]$ of the
operators above.  Recall that the {\em formal adjoint} of a
differential operator between vector bundles, $\cD:E\to F$, is a
differential operator $\cD^*:F^*\to E^*$, provided a smooth measure is
fixed.  Given a metric $ g$ from the conformal class we have the
Riemannian measure.  This depends on the choice $ g$. However there is
a canonical conformal volume form $ \vol$, that is the canonical
section of $ \ce_{[a_1a_2\cdots a_n]}[n]$ compatible with the
conformal metric. Thus we can invariantly integrate densities of
weight $ -n$.  As a result, the formal adjoint, computed with respect
to conformal structure, of a conformally invariant differential
operator $\cD:\cE^s_t[w]\to\cE^u_v[w']$, where $s,t,u,v$ are index
arrays rather than single indices, will be a {\it conformally}
{\it invariant} differential operator
\begin{equation}\label{genlFA}
\bar{\cD}:\cE^v_u[-n-w']\to\cE^t_s[-n-w].
\end{equation}

Setting $\ulw=1-w-n$, the formal adjoint of $\S_A{}^{\bar{A}} $ is
easily found (integrating by parts) to be the operator
$\bar{\S}^{\bar{A}}{}_A:\ce^A[\ulw]\to \ce^{\bar{A}}[\ulw]$ given by
$$
[ \bar{\S}^{\bar{A}}{}_A V^A]_g=
\left(\begin{array}{c} \mu^a -\dfrac1{\ulw+1}\nd^a\si \\
                       \dfrac1{n+2\ulw}\left(
-\dfrac1{\ulw+1}\Delta+\J\right)\si
+\tau
\end{array}\right)
$$
if $[V^A]_g=(\si,\mu^a,\tau)$.
In the alternative notation,
$$
\begin{array}{l}
\bar{\S}^{\bar{A}}{}_AV^A=Z^{\bar{A}}{}_aZ_A{}^aV^A
-\dfrac1{\ulw+1}Z^{\bar{A}}{}_a\nd^a(X_AV^A) \\
{}\qquad
+\dfrac1{n+2\ulw}X^{\bar{A}}\left(-\dfrac1{\ulw+1}\Delta+\J\right)(X_AV^A)
+X^{\bar{A}}Y_AV^A.
\end{array}
$$
It follows from the splitting property of $ \S$ that $ \bar{\S}$ splits
the canonical bundle injection $\ce^{\bar{A}}[-3]\to
\ce_A[-3]$. That is, upon restriction to $\ce^{\bar{A}}[-3]$,
regarded as a subbundle of $\ce_A[-3]$, the operator $ \bar{\S}$ is the
identity.

The formal adjoint
$\bar{I}^a{}_{\bar{A}}$ of
$I_{\bar{A}}{}^a$ is the map which  simply takes
$\ce^{\bar{A}}[\ulw]=\ce_a[\ulw+1]\lpl\ce[\ulw-1]$
to its quotient
by the subbundle $
\ce[\ulw-1]$, so if $V^{\bar{A}}\in\ce^{\bar{A}}[\ulw] $ is given by $
[(\mu^a,\tau)]_g$, then $[\bar{I}^a{}_{\bar{A}}V^{\bar{A}}]_g=\mu^a$.

To construct a gauge extension of the Maxwell operator we merely
have to specialise to $n=4 ,w=0$ and compose with $\Box$
appropriately. We obtain the following.

\begin{theorem}\label{Maxcurved}The operator
$$
(E^{{\bar{A}}a}:=\bar{\S}^{\bar{A}A} \Box \S_A{}^{\bar{C}}I_{\bar{C}}{}^a):
\ce_a \to \ce^{\bar{A}}[-3]
$$
is a conformally invariant operator such that\newline
{\rm(i)} $\bar{I}_{b\bar{A}}E^{{\bar{A}}a}:
\ce_a \to \ce_b[-2]$
is a non-zero multiple of the Maxwell operator.\\
{\rm(ii)} $-2E^{\bar{A}a} \nd_a f=X^{\bar{A}}P_4 f$\\
{\rm(iii)} $ E$ is elliptically coercive.
\end{theorem}

To adapt figure \ref{opext} to this setting, we have that $ \cA$
is $ d$ on functions, and $ \cB$ is the Maxwell operator $ \d d$. Then (i)
is stating that $E $ is an extension (cf.\ $ \cL$) of the Maxwell
operator; (ii) means it is a gauge extension (since the Paneitz
operator is elliptic), and
(iii) says finally that it is an elliptically coercive
gauge extension.

\vspace{.4cm}

\noindent{\bf Proof of the theorem:} The formulae are
conformally invariant by construction.  To obtain (i) the key point is
to establish that $ E$ is non-trivial.
Since $\Box$ is elliptic  (see section \ref{conf}) it has
finite dimensional null space in any compact setting. Thus it follows
that the composition $ \Box S I: \ce_a\to \ce_{A}[-3]$ is not trivial.
Now $ \ce_A[-3] = \ce[-2]\lpl \ce_a[-2]\lpl \ce[-4]$. Composing with $
\Box S I$ the map $\ce_A[-3] \to \ce[-2] $ yields an invariant
differential operator $ \ce_a\to \ce[-2]$. Consulting the diagram
(\ref{maxbgg}) we see that there is no such operator in the flat
model. Thus in that homogeneous setting this last operator must be
trivial, meaning that the image of $ \Box S I$ lies in the sub-bundle $
\ce^{\bar{A}}[-3]=\ce_a[-2]\lpl \ce[-4] $ of $ \ce^A[-3]$. It follows
easily that on the conformal sphere $\overline{S}$ acts as the
identity on the image of $ \Box S I$. Thus $E=\bar{S} \Box S I $
is also non-trivial in the flat model and so non-trivial in general.

Now suppose $ \bar{I} E$ were trivial on the
conformal sphere. Then, since $\ce^{\bar{A}}[-3]$ has the composition
series $\ce^{\bar{A}}[-3]=\ce_a[-2]\lpl \ce[-4] $, $ E$ would give a
non-trivial invariant operator $\ce_a\to \ce[-4]$. But according to
the diagram above there is no such operator. Thus $ \bar{I}
E$ is non-trivial in this flat model and hence non-trivial in general.
Once again from the diagram it follows that on the conformal sphere $
\bar{I} E$ is the Maxwell operator (up to a non-zero scale). In
fact it is easily verified that even in the general case the Maxwell
operator is the unique conformally invariant differential operator
between the bundles $ \ce_a$ and $ \ce_a[-2]$.  This concludes the
proof of (i).

Now since exact 1-forms are annihilated by the Maxwell operator it
follows from (i) that $ Edf$ takes values in the subspace $ X \ce[-4]$
in $\ce^{\bar{A}}[-3]=\ce_a[-2]\lpl \ce[-4]$. That is
$$
E^{\bar{A}a}\nd_a f= X^{\bar{A}}P f
$$
for some invariant operator $ P:\ce\to \ce[-4]$. By construction $
P$ factors through $ d$ and so annihilates constant functions. Using
once again the ellipticity of $ \Box$ we can also deduce that $ P$ is
non-trivial and so by uniqueness $ P$ is the Paneitz operator as
claimed in (ii).

Finally observe that it is straightforward to directly calculate the
operators in the proposition. Choosing some metric $ g$ from the conformal class for the purpose of calculations, observe that
$(n+w-2)\S_A{}^{\bar{A}}I_{\bar{A}}{}^a: \ce_a[w]\to \ce_A[w-1]$ is
simply $\phi_a \to (0, ~(n+w-2)\phi_a,~-\nabla^b\phi_b)$ (this is the
operator $ E^{bC}$ of \cite{MEsrni}).
Setting $ n=4,w=0$ and composing with $ \Box$ yields
$$
\Box \S_A{}^{\bar{C}}I_{\bar{C}}{}^a \phi_a =\left(\begin{array}{c} 0 \\
                       2 \nd^b\nd_{[b}\phi_{a]}\\
                       -\frac12\nd_b(\nd^b\nd^c+4\V^{bc}-2\J\bg^{bc})\phi_c
\end{array}\right)
$$
Thus
$\bar{\S}^{\bar{A}B} \Box \S_B{}^{\bar{C}}I_{\bar{C}}{}^a\f_a$ is just
$$
\left(2 \nd^b\nd_{[b}\phi_{a]}\,,
-\tfrac12\nd_b(\nd^b\nd^c+4\V^{bc}-2\J\bg^{bc})\phi_c\right).
$$
So we have that $[E\phi]_g = -(\d d\phi, \d d \d \phi + \LOT )$.
As mentioned already, the elliptic coercivity of this is verified by
composing with $(\d d, d) $ which yields $\D^2+\LOT$.  \quad $ \Box$

\vspace{.4cm}

Some points are worth making here. Firstly note that the second
component, $ \sS \phi:=
-\tfrac12\nd_b(\nd^b\nd^c+4\V^{bc}-2\J\bg^{bc})\phi_c$, of $ [E]_g$ is
Eastwood and Singer's gauge operator, at least modulo a factor of $
-1/2$. From this explicit formula we see that we have $ \sS =\d T$ for
a second order operator $ T$, a fact that we will use below.  Next
from its construction here, we see that the conditional conformal
invariance of the gauge operator $ \sS$ (i.e.\
the fact that it is conformally
invariant on solutions of the Maxwell operator) is an immediate
consequence of the invariance of the operator $ E^{\bar{A} a}$ and the
conformal transformation law $[V^{\bar{A}}]_g= (\mu_b, \tau)\mapsto
\widehat{(\mu_a,\tau-\Upsilon^c\mu_c)} =[V^{\bar{A}}]_{\hat{g}}$ for
sections $ V^{\bar{A}}$ of $\ce^{\bar{A}}[-3]$.  Finally we should say
that, although the operator $\bar{\S}^{\bar{A}B}$ effectively plays no
role here, we can only know this by actually performing the
calculation in some detail.

\subsection{Application: A conformal Hodge theory} \label{hodge}

Here we suppose that $ M$ is an oriented
compact 4-manifold. If $M$ is equipped with a Riemannian metric, then
Hodge-de Rham theory identifies the $ i^{\rm th}$ de Rham cohomology $
H^i(M)$ with the space of harmonics $ \cH^i(M)$. This is the kernel of
the form Laplacian $\d d +d\d $ on $ i$-forms or, alternatively, it is
recovered by $ \cH^i(M)=\ker(d:\ce^i\to\ce^{i+1})\cap \ker(\d : \ce^i
\to \ce^{i-1})$ (with obvious qualifications at either extreme of the
de Rham complex).  As before, $\ce^i:=\ce_{[a_1\cdots a_i]}\,$.

In general then we would expect the subspace of harmonics to move
around as we change to different metrics in the conformal class.  In
fact in dimension 4, $ \cH^2(M)$ is a conformally invariant subspace
of $\ce^2$. This is obvious as both $ d$ and $ \d$ are conformally
invariant on $ \ce^2$. Also $\cH^0(M)$ is just the invariant subspace
of locally constant functions. On the other hand $ \cH^1(M)$ is not
stable in this way in $ \ce^1$.  Verifying this is the same
calculation as verifying the failure of the Coulomb gauge to be
conformally invariant (see Section \ref{conf}).  It is interesting to
ask whether there is a conformally invariant replacement for $\cH^1(M) $.
In fact $\ker( E)$ is, at the very least, a good candidate.
\begin{theorem}\label{confhodge}
Suppose $ M$ is an oriented compact manifold such that the null space of
$P_4$ is the space of locally constant functions. Then
$\ker(E:\ce_a\to \ce^{\bar{A}}[-3])$ is a
  conformally invariant subspace of $\ce^1 $ isomorphic to $H^1(M)$.
\end{theorem}
Note that since, in a choice of scale, $[E\phi]_g $ has the form $
(-\d d \phi,\sS \phi )$ it follows that $ \ker(E)$ is just $ \ker(\d
d: \ce^1\to \ce^1[-2]) \cap \ker(\sS:\ce^1\to \ce[-4])$. This
intersection is conformally invariant because $ \sS$ is invariant on $
\ker(\sS)$.

Note that for $i=0,1,2 $ there is an invariant pairing between $ \ce^i$ and
$ \ce^{4-i}$ given simply by
\begin{equation}\label{invtpair}
\phi,\psi \mapsto \int_M \phi\wedge \psi
\end{equation}
Note this does not require a Riemannian or even a conformal structure;
it is well defined on any oriented
compact 4-manifold $M$. So of course, in particular, it is conformally
invariant.

At a Riemannian scale, the Hodge $\star$ operator
$\f_{a_1\cdots a_k}\mapsto\epsilon^{c_1\cdots c_k}{}_{b_1\cdots b_{n-k}}
\f_{c_1\cdots c_k}$ is a natural bundle
isometry $\ce^k\to\ce^{n-k}$ in the form
inner products $f_k(\f,\psi):=
(k!)^{-1}\f_{a_1\cdots a_k}\psi^{a_1\cdots a_k}\,$, and we have
the identity $f_k(\f,\psi)\epsilon=\f\wedge\star\psi$.
Given just a conformal structure, $\star$ carries $\ce^k[w]$ to
$\ce^{n-k}[w+n-2k]$.  We can use this to rewrite the total
space of the de Rham complex (now in dimension 4) as
$$
\ce^*: =\ce\oplus \ce^1 \oplus \ce^2 \oplus \ce^1[-2] \oplus \ce[-4].
$$
The invariant pairing \nn{invtpair} then gives a conformally
invariant non-degenerate, indefinite inner product on the vector space
$ \ce^*$. This is determined by symmetry, bilinearity and the formulae
$$
(\phi,\psi)=\left\{\begin{array}{ll}
                                   \int_M \phi\wedge\star
\psi,\  &  \phi\in \ce^k, \psi\in \ce^{k}[2k-4],\ k=0,1~{\rm or}~2,\\
                           0  & {\rm otherwise}.
\end{array} \right.
$$
The non-degeneracy of this follows directly from the positive
definiteness of each $f_k$.  Note also that for $ \phi,\psi\in \ce^2$, we have
$\int\f\wedge\star\psi = \int\psi\wedge\star\f$ since in either case
the integrand is $f_2(\f,\psi) \epsilon$. Thus there is no conflict
with the extension by symmetry.
(The restriction to compact $ M$ can be lifted by requiring
at least one form in the inner product to have compact support,
and of course the generalisation to other
dimensions is straightforward.)

Invariant operators on a subspace of $\ce^*$ will be identified with
their trivial extension to an operator on $\ce^*$.  In this way we
define the formal adjoint of operators between subspaces of $ \ce^*$.
For example the formal adjoint of $d:\ce\to\ce^1$ is the conformally
invariant
operator
$\d:\ce^1[-2]\to \ce[-4]$.  In this picture the
formal self-adjointness of $ P_4$ means that $ (\phi,P_4
\psi)=(P_4 \phi, \psi)$ for any $ \phi,\psi\in \ce^*$. Of course the
only real content of this is just that $ (f,P_4 h)=(P_4 f, h)$ for any
$ f,h\in \ce$.

\vspace{.4cm}

\noindent{\bf Proof of the theorem:} First note that if, for $ \F\in \ce^1$,
$ E\F=0$ then clearly $\d d \F =\bar{I} E \F=0$. So $ 0=(\F,\d d
\F)= (d \F,d \F)$ and hence $ d\F=0$, since on 2-forms our inner
product agrees with the usual form inner product.
So there is a map from $ \ker(E)$ to $ H^1(M)$ given by mapping the
closed form $ \F$ to its class $ [\F]$ in $ H^1(M)$.

On the other hand any closed 1-form $ \F$ satisfies $ \bar{I} E
\F=0$.  (So $ \bar{I} E \F=0$ is equivalent to $ \F$ being
closed.) To obtain a map from $ H^1(M)$ to $\ker(E)$ it remains to
verify that there is a unique element $ \F'$ in the class $ [\F]$
satisfying $\sS \F'=0$. Note that $ \F'=\F+df $ for some $ f\in \ce^0$.
So this equation is
\begin{equation}\label{findrep}
\sS \F +\sS d f =0 .
\end{equation}
That is $2 \sS \F =P_4 f$ since, by (ii) of theorem \ref{Maxcurved}, $ \sS df =-\frac{1}{2}P_4 f$ for any function $ f$.

Now since $ P_4$ is elliptic, formally self-adjoint and has $
\ker(P_4)\subseteq \ker(d) $ (i.e. in the terminology of section
\ref{extend}, $ P_4$ has harmless kernel) it follows that $\image( P_4
)$ is precisely the subspace of $\ce[-4]$ orthogonal to the space of
locally constant functions (recall $\ce[-4]$ pairs with $ \ce$).
Recall that, in any choice of conformal scale, $ \sS$ is of the form $
\d T$ for a second order operator $ T$ on forms.  Thus $\sS \F $ lies
in the subspace of $ \ce[-4]$ orthogonal to locally constant
functions.  That is in $\image( P_4) $.  (Note that for $ h\in \ce$
and $ \phi \in \ce^1$ we have, in any choice of conformal scale,
$(h,\sS \phi)=(h,\d T \phi)=(d h, T\phi)$.
Although $ T\phi$ is not conformally invariant $ (d h, T\phi)$ is
conformally invariant.) Thus there is a unique $ df$ solving
\nn{findrep}. So for any closed 1-form $ \F$ there is a unique element
$\F' $ in the class $ [\F]$ satisfying $ E\F'=0$ and this gives a well
defined map $ H^1(M)\to \ker(E)$. This clearly inverts the map from
$ \ker(E)$ to $ H^1(M)$ described above. \quad $\Box$

\section{The deformation complex} \label{defcx}

There is an important analogue of the above construction which we
believe will have a significant role in the deformation theory of
conformal structures.  On any manifold let us write $\cR_{ab}{}^c{}_d$
or simply $ \cR$ to denote the bundle of tensors with the same
algebraic symmetries as the Riemann tensor of an affine connection.
Now consider the second order universal operator mapping
metrics to their Weyl curvature tensors, $ g\mapsto C(g)$. Linearising about
a given metric $ g$ leads to an operator on $\ce_{(ab)}$ taking values
in the bundle $\cR$.  Notice that the conformal invariance of the Weyl
tensor formula means that this operator  annihilates the trace part
of perturbations, and so the restriction to $ \ce_{(ab)_0}$ fully
captures the operator and also means that the operator yields a well
defined operator linearising perturbations of conformal structures.
Viewing this as an operator on perturbations of a given conformal
metric $ \bg$ gives the operator we will denote $ \cD_1:
\ce_{(ab)_0}[2]\to \cR$. Of course the image lies in the totally trace-free
(with respect to $ \bg$) part of $ \cR$, which we shall
call $\cW$ or
$\cW_{ab}{}^c{}_d$.

It is well known
that, on conformally flat
structures, the local kernel of $ \cD_1$ is the image of the (conformally
invariant) conformal Killing operator $ \cD_0: \ce^a\to \ce_{(ab)_0}[2]$
which is given by $ t^a\mapsto \nd_{(a}t_{b)_0} $. We may regard $ h$ in
$\ce_{(ab)_0}[2] $ as a potential for the linearised curvature $ \cD_1 h$
and the transformations $ h\mapsto h+ \cD_0 t$ as gauge freedom.

Returning to the general setting note that the bundle $ \cW$ splits
into self-dual and anti-self-dual components that we denote $ \cW^+$
and $ \cW^-$ respectively. Composing these projections with $\cD_1$
yields operators $ \cD_1^+$ and $ \cD_1^-$. We will construct further
operators as formal adjoints of these.  Let $ \cW^*$ be the direct sum
space,
$$
\cW^*:= \ce^a\oplus \ce_{(ab)_0}[2]\oplus \cW \oplus
\ce_{(ab)_0}[-2] \oplus \ce_a[-4].
$$
Note that a section $ t^a\in \ce^a$ can be paired in a conformally
invariant way with $ w_a\in \ce_a[-4]$, $(t,w):=\int_M t^aw_a$.
Similarly we have $(h,B)=\int_M h_{ac}B^{ac}$ for $h_{ac}\in
\ce_{(ac)_0}[2],~ B_{ac}\in \ce_{(ac)_0}[-2]$ and for $ U,V\in \cW$
there is $ \int_M U_{abcd}V^{abcd}$. Setting all other pairings
between direct sum components of $ \cW^*$ to be zero and requiring
bilinearity determines a conformally invariant indefinite (but
non-degenerate) inner product on $ \cW^*$ similar to the one on $
\ce^*$. Also similar to that case, we identify operators on components or
subspaces of $ \cW^*$ with their trivial extension to operators on $
\cW^*$.  It follows immediately that any conformally invariant
operator between components of $ \cW^*$ has a formal adjoint and this
is another conformally invariant operator. In particular we have the
formal adjoints: $\bar{\cD_1}$ of $ \cD_1$, $ \bar{\cD_1}^{\pm}$ of $
\cD_1^{\pm}$ and $\bar{\cD_0}$ of $ \cD_0$.  Here by $\cD_0 $ we mean
the conformal Killing operator in the general (conformally curved)
setting; this is given
by the same formula as above.

Summarising the situation we have the sequence and operators indicated
by the solid arrows in the following diagram.
\begin{equation}\label{defbgg}
\setlength{\unitlength}{.95pt}
\begin{picture}(300,120)(58,-55)
\put(49,-20){\line(0,-1){30}}
\put(49,-50){\line(1,0){15}}
\put(80,-50){\line(1,0){15}}
\put(110,-50){\line(1,0){15}}
\put(140,-50){\line(1,0){15}}
\put(170,-50){\line(1,0){15}}
\put(200,-50){\line(1,0){15}}
\put(230,-50){\line(1,0){15}}
\put(260,-50){\line(1,0){15}}
\put(290,-50){\line(1,0){15}}
\put(320,-50){\line(1,0){15}}
\put(340,-50){\line(1,0){4}}
\put(344,-50){\vector(0,1){30}}
\put(43,0){$\ce^a$}
\put(90,0){$\ce_{(ab)_0}$} 
\put(100,16){\line(0,1){36}}
\put(100,52){\line(1,0){179}}
\put(279,52){\vector(0,-1){37}}
\put(184,28){$\cW^{+}$}  
\put(184,-32){$\cW^{-}$} 
\put(255,0){$\ce_{(ab)_0}[-2]$} 
\put(330,0){$\ce_a[-4]$}    
\put(63,4){\vector(1,0){15}} 
\put(140,15){\vector(1,1){15}}     
\put(140,-15){\vector(1,-1){15}}   
\put(235,29){\vector(1,-1){15}}    
\put(235,-29){\vector(1,1){15}}    
\put(310,4){\vector(1,0){15}}      
\end{picture}
\setlength{\unitlength}{1pt}
\end{equation}
The operator $ \ce_{(ab)_0}[2]\to \ce_{(ab)_0}[-2]$ is defined here to
be $ \bar{\cD_1}\cD_1$.  Now on the conformal sphere case some now
well known representation theory can produce a similar sequence of
operators -- see for example \cite{ESlo}.  From that theory we also
know that in that setting of the flat model several things are true:
There is also a conformally invariant operator $L:\ce^a\to
\ce_a[-4]$ as indicated by the long arrow in the diagram. All the
operators in the diagram are unique (up to scale) and the diagram
(including $ L$) gives a complete set of the conformally invariant
differential operators
with the
bundles concerned as domain or range bundles.
In fact the existence
of the differential operators is purely a local issue and these
results all carry over to the general
conformally flat setting. Let us write $ B$ for the operator $
\bar{\cD_1}\cD_1$ on a conformally flat compact manifold.  Note that
the formal adjoint of $ L$ is also a non-trivial conformally invariant
operator $\ce^a\to \ce_a[-4] $. Thus by uniqueness $ L$ is
formally self-adjoint.

We are now in a similar setting to the Maxwell problem considered
above.  In fact the situation here is still somewhat more complicated
and this affects the overall progress we will make below. What we will
show is how to construct a formally-self-adjoint conformally invariant
curved analogue of $ B$ and a conformally invariant elliptically coercive
extension of this. Then in the conformally flat case
we will use this to isolate the subspace of $ \ce_{(ab)_0}[2]$
corresponding to the first cohomology of the complex $ \ce^a\to
\ce_{(ab)_0}\to \cW$.

A well known conformal invariant is the Bach tensor. In terms of the
Weyl tensor this is the trace-free symmetric tensor of weight $ -2$
given by $ B_{ab}:= \nd^c\nd^dC_{acbd}+\Rho^{cd}C_{acbd}$. Arguing as
above it is straightforward to conclude that the linearisation of this
is a conformally invariant operator on $ \ce_{(ab)_0}[2]$. Since $
B_{ab}$ is trace-free it follows that perturbations on a conformal
manifold with vanishing Bach tensor yield an operator
$\ce_{(ab)_0}[2]\to \ce_{(ab)_0}[-2] $. Clearly then in the
conformally flat case $ B$ agrees with this linearisation of the Bach
tensor and so we will refer to $ B$ as the Bach operator.

\subsection{The extended Bach operator}

Let us once again return to arbitrary dimension $ n\geq 3$.  {} Recall
that $\ce_{\bar{A}}[-1]$ has the composition series $
\ce_{\bar{A}}[-1]= \ce\lpl\ce_a$. So tensoring with $ \ce_b$ we have $
\ce_{\bar{A}b}[-1]=\ce_b\lpl \ce_{ab}$. Thus there is a canonical
bundle injection $\ce_{[ab]}\oplus \ce[2]\to \ce_{\bar{A}b}[-1]$ and
we define $ \cF_{\bar{A}b}[-1]$ to be the quotient. Tensoring now with
$ \ce[w+2]$ gives $ \cF_{\bar{A}b}[w+1]$ which has the composition
series
$$
\cF_{\bar{A}b}[w+1] =\ce_{b}[w+2]\lpl \ce_{(ab)_0}[w+2].
$$
We define $
\BI_{\bar{A}}{}^a: \ce_{(ab)_0}[w+2]\to \cF_{\bar{A}b}[w+1]$ to be the
obvious inclusion. Thus, for example,
for $ h_{ab}\in\ce_{(ab)_0}[w+2] $, we have
$[\BI_{\bar{A}}{}^a h_{ab}]_g =(0,h_{ab})$.
In the alternative notation,
$$
\BI_{\bar{A}}{}^ah_{ab}=Z_{\bar{A}}{}^ah_{ab}\,.
$$

Clearly there is a bundle surjection $
\ce_{Ab}[w+1]\to\cF_{\bar{A}b}[w+1] $.  For $ w\neq -2,-n,-n/2$ there
differential splitting operator (cf.\ \nn{diffsplit1}) $
\BD_B{}^{\bar{B}}: \cF_{\bar{B}a}[w+1] \to \cE_{Ba}[w+1]$ given by
$( Y_{\bar{B}}\s_a+Z_{\bar{B}}{}^b s_{ba})
\mapsto (Y_B\s_a+Z_B{}^b S_{ba}+X_B\rho_a $)
where
$$
S_{ba}=s_{ba}+ \frac{\nd_{[b} \s_{a]}}{w+2} + \frac{\bg_{ba}\nd^c\s_c}{n(n+w)}
$$
and
$$
\begin{array}{l}
\rho_a=-\dfrac{1}{n+w}\left[\nd^b s_{ba}+\tfrac{1}{2}\J\s_a+\dfrac{n-2}{2n}
\V_a{}^b\s_b\right] \\
{}\qquad+\dfrac{n-2}{2(n+2w)(n+w)(w+2)}\Bigg[\dfrac{(n+2w+4)}{n}
(\nd_a\nd^c\s_c+(n+w)\V_a{}^b\s_b) \\
{}\qquad-(\nd^c\nd_c \s_a+(w+1)\J \s_a)\Bigg] .
\end{array}
$$

Both of these operators exist and the corresponding formulae are valid
if the fields concerned take values in other tractor bundles. This is
trivial for $\BI_{\bar{A}}{}^a $ and straightforward to verify for
$\BD_B{}^{\bar{B}}$. We do not need this here for these operators, but
we shall for their formal adjoints, for which it follows
automatically. The formal adjoint of $\BD_B{}^{\bar{B}}$, for example,
yields a conformally invariant operator $\bar{\BD}^{\bar{C}B}
:\cE^\Psi_{Ba}[1+\ulw]\to \bar{\cF}_{\Psi a}^{\bar{C}}[1+\ulw]$ where
$\ulw=-n-w$ and $\ce^\Psi=\ce_\Psi$ indicates any tractor bundle
(tensored here into $ \cE_{Ba}[1+\ulw]$ and
$\bar{\cF}^{\bar{C}}{}_{a}[1+\ulw]$).  It is straightforward to
calculate explicit formulae for the formal adjoints.

Finally, we need the operator
$$
\AD^{\b l}:\cE^\Psi_{l}[w+1]\to \cE_\Psi^\b [w-1]
$$
and its formal adjoint, where again $\ce^\Psi=\ce_\Psi$ indicates
any tractor bundle. Omitting the $ \Psi$, the operator sends $h_{l}$
to $\AD^{\b l}h_{l}\,$, where
$$
[\AD^{\b l}h_{l}]_g=\AY^\b{}_lh^l
+\AZ^\b{}_k{}^l\m_{l}{}^k+\bW^\b\phi+\AX^{\b k}\r_{k}\,.
$$
Here
$$
\begin{array}{rl}
\phi&=\dfrac1{n+w-1}\N_kh^k\,, \\
\m_{kj}&=\dfrac1{w+1}(\N_kh_{j}-\N_jh_{k}), \\
\r_{j}&=\dfrac1{(w+1)(n+2w-2)}\Bigg[\N^l\N_lh_{j}+(w)\J h_{j} \\
&\qquad -\dfrac{n+2w}{n+w-1}(\N_j\N^qh_{q}+(n+w-1)\V_j{}^qh_{q})
\Bigg].
\end{array}
$$
%\edz{NB: This operator appears to be unique -- i.e. no way to add
%  Weyl curvature.}
The formal adjoint of this,
expressed in terms of $\ulw=-n-w$, is an operator
$$
\bar{\AD}_{\b a} :\cE^\b[\ulw+1] \to \cE_a[\ulw+1] .
$$

Setting $ $n=4 and $ w=0$ we obtain the following:

\begin{theorem}\label{Bcurved}The operator
$$\textstyle
(F^{\bar{B}}{}_{d}{}^{ab}:=\bar{\BD}^{\bar{B}A} \bar{\AD}_{d\b}
\Box\AD^{\b b}\BD_A{}^{\bar{A}}\BI_{\bar{A}}{}^a): \ce_{(ab)_0}[2]\to
\cF^{\bar{B}}{}_{d}[-3]
$$
is a  conformally invariant elliptically coercive operator such that
$$\textstyle
({\bB}_{cd}{}^{ab}:=\bar{\BI}_{c\bar{B}}\bar{\BD}^{\bar{B}A} \bar{\AD}_{d\b}
\Box \AD^{\b b}\BD_A{}^{\bar{A}}\BI_{\bar{A}}{}^a): \ce_{(ab)_0}[2]\to
\ce_{(cd)_0}[-2]
$$
is a formally self-adjoint curved analogue  of the  operator $ B$.
\end{theorem}
\noindent{\bf Proof:} Conformal invariance is clear and the
last displayed operator is formally self-adjoint by construction.  The
final claim follows from the uniqueness of $ B$ and an argument which
completely parallels the corresponding point in the Maxwell case.  It
remains to establish elliptic coercivity.

We will show the operator $ F$, when evaluated in any scale, is a factor
of that scale's $(\nd^a\nd_a)^4$, modulo lower order terms; this will
establish elliptic coercivity.  Suppose
a scale is chosen and we decompose the bundles in the usual way; then
direct computation with the above formulas yields the principal parts
$$
\begin{array}{l}
(\BI F= {\bB}):h_{ab}\mapsto H_{ab}:=
-\tfrac12\nd^j\nd_j\nd^i\nd_ih_{ab}
+\nd^j\nd_j\nd^i\nd_{(a}h_{b)i} \\
{}\ \ -\tfrac13\nd_a\nd_b\nd^j\nd^ih_{ij}
-\tfrac16\nd^k\nd_k\nd^j\nd^ih_{ij}g_{ab}
\end{array}
$$
and
$$
{\gG}: h_{ab}\mapsto\eta_a:=-\tfrac16\nd_a\nd^k\nd_k\nd^j\nd^ih_{ij}
+\tfrac18\nd^k\nd_k\nd^j\nd_j\nd^ih_{ia}
$$
for the operator and gauge-part
 respectively.  A useful way to express
these principal parts is in terms of the conformal Killing operator
$\K$ given above
(using the scale to identify vector fields and one-forms):
$$
\begin{array}{rl}
{\bB}&=-\tfrac13(\K \K^*+\D)(\K\K^*+\tfrac32\D)+\LOT, \\
{\gG}&=\K^*\left(\tfrac7{18}(\K\K^*)^2+\tfrac{47}{36}\K\K^*\D+\tfrac{25}{24}\D^2
+\LOT\right).
\end{array}
$$
(Here we are using $\K^*$ for the formal adjoint of $ \K$.
Elsewhere we have used overbars to indicate formal adjoints. The point is here we mean this in the usual Riemannian sense and we want to distinguish this from the conformally invariant $ \bar{\cD}_0$.)
The
elliptic deficiency in $ \bB$ is clear from the fact that the sixth-order
symbol of the operator
$$
(\K\K^*+\D)(\K\K^*+\tfrac32\D)\K\K^*
$$
vanishes.  This shows that the leading symbol of
$(\K\K^*+\D)(\K\K^*+\tfrac32\D)$ annihilates the (non-trivial)
range of the leading symbol of $\K\K^*$, and so cannot be invertible.
On the other hand,
$$
\horisys{a_1(\K\K^*)^2+a_2\K\K^*\D+a_3\D^2}{\K(a_4\K\K^*+a_5\D)}
\vertsys{{\bB}}{{\gG}}=\D^4+\LOT
$$
for $(a_1,a_2,a_3,a_4,a_5)=-(56,\frac{340}3,2,48,56)$.\quad$\Box$

\begin{remark} \label{Ppart}
  {\rm When $ \image(\K)\subset \ker({\bB})$ (as, for example, in the
    next section), the operator $ {\gG} \K$ will be conformally
    invariant, so it is of interest to examine it more closely.
    In general it
    has the form
$$
\begin{array}{rl}
{\gG}\,\K
&=\K^*\left(\tfrac7{18}(\K\K^*)^2+\tfrac{47}{36}\K\K^*\D+\tfrac{25}{24}\D^2
+\LOT\right)\K \\
&=\tfrac7{18}(\K^*\K)^3+\tfrac{47}{36}(\K^*\K)^2\D+\tfrac{25}{24}\K^*\K\D^2
+\LOT.
\end{array}
$$
Since
$$
\K^*\K=\d d+2\tfrac{n-1}{n}d\d+\LOT
$$
in dimension $n$, we have that
\begin{equation}\label{GgS}
\begin{array}{rl}
{\gG}\,\K&=\frac18(\d d)^3-\frac1{16}(d\d)^3+\LOT \\
&=(\frac18\d d-\frac1{16}d\d)\D^2+\LOT
\qquad(n=4).
\end{array}
\end{equation}
These coefficients check with \cite{tbjfa}, Remark 3.30, which
shows that an order $2p$
invariant operator $\cE_{[a_1\cdots a_k]}[w]\to\cE_{[a_1\cdots a_k]}[w']$
in the conformally flat case must have $w=-(n-2k-2p)/2$ and
$w'=-(n-2k+2p)/2$, and must take the form $w'(\d d)^p+w(d\d)^p+\LOT$
up to a constant factor.
The operator in \nn{GgS} is elliptic, though not positively so.  That is,
its leading symbol is invertible but not positive definite.  To check
invertibility, just note that if
$a\ne 0\ne b$, then $(a^{-1}\d d+b^{-1}d\d)(a\d d+bd\d)=\D^2+\LOT$.
In particular,
$$
(8\d d-16d\d){\gG}\,\K=\D^4+\LOT.
$$
}\end{remark}

\subsection{Application: The moduli space of conformally flat deformations}

Recall that on a conformally flat manifold $ B$ is the operator $
\bar{\cD_1}\cD_1$.  It follows from the uniqueness of the operators in
the pattern \nn{defbgg} that $B$ is twice the composition around
either edge of the diamond. Thus $ (\bar{\cD_1}^+,-\bar{\cD_1}^-)$
annihilates the image of $ (\cD_1^+,\cD_1^-)$ and so there is a
conformally invariant resolution
$$
0\to\bW\to \ce^a\to \ce_{(ab)_0}[2]\to \cW \to \ce_{(ab)_0}[-2]\to \ce_a[-4]\to 0
$$
where $\bW$ is the space of conformal Killing vectors.  We will
write $ \cD_2$ for the operator $ \cW \to \ce_{(ab)_0}[-2]$ and the
last operator is just $\bar{ \cD}_0$, the formal adjoint of $ \cD_0$.
 (In fact on
conformally flat structures conformal Killing vectors correspond to
parallel adjoint tractors and for the conformal sphere case $\bW$
is isomorphic to $ {\mathfrak{so}}(n+1,1)$. This will be discussed elsewhere
\cite{GoCapinprogress}.)

Since by definition $\bar{\cD_1}$ is the formal adjoint of $ \cD_1$
the second cohomology of the resolution is, according to standard
Hodge theory, isomorphic to $ \ker(\cD_2)\cap \ker (\bar{\cD_1})$,
that is the space of harmonics in $ \cW$. As the operators $ \cD_i$
are conformally invariant this subspace is conformally invariant.
This is analogous to the de Rham setting above. Also in a parallel to
that case we will see that the gauge extension of the Bach operator is
related to the first cohomology. First we note that the latter has nice
interpretation.

We observed already that $ \ce_{(ab)_0}[2]$ is the space of
infinitesimal conformal metric deformations. Thus the kernel of the
map $ \cD_1:\ce_{(ab)_0}[2]\to \cW$ consists of deformations preserving
conformal flatness.  On the other hand the image of the conformal
Killing operator $\cD_0: \ce^a\to \ce_{(ab)_0}[2]$ is the subspace of
deformations coming from infinitesimal diffeomorphisms. Thus the first
cohomology of the complex is the formal tangent space to the moduli
space of conformally flat structures. Toward the question of
integrability of deformations, Calderbank and Diemer \cite{calderTD}
have shown that if the second cohomology vanishes then all
deformations can be formally integrated (to a power series).

Before we prove the main result let us observe that we are once again
fully in the setting of figure \ref{opext}. By the uniqueness of $ B$
it is clear that it is recovered, in the conformally flat setting, by
the operator ${\bB}=\bar{\BI}F$ from theorem \ref{Bcurved}. Now since
$ \cD_1\cD_0=0$ and $ B=\bar{\cD_1}\cD_1$ it is immediate that $
B\cD_0=0$.  Thus the theorem gives $ F$ as an elliptically coercive
gauge extension of $ B$. Then $ {\gG}$ is a corresponding gauge
operator in the sense of figure \ref{opext}. Since the image of $
\cD_0$ is in the null space of $ B$ it follows that $
{\gG}\cD_0:\ce^a\to \ce_{a}[-4]$ is conformally invariant. Using the
ellipticity of $ \Box$ and arguments similar to those used in the
proof of theorem \ref{Maxcurved} it is clear that this is non-trivial.
In fact we have already verified explicitly in remark \ref{Ppart} that
this is elliptic.  So up to scale ${\gG}\cD_0 $ must agree with $L$.
Let us henceforth take $ L$ to be this
elliptic operator. By construction here we have that $ L$ factors through $
\cD_0$. On the other hand we have already observed that it is
formally-self-adjoint. Thus $ L= \bar{\cD_0}U\cD_0$ for some operator
$ U$.  It follows easily
that on $\ker( B) $, ${\gG} $ has the form $ \bar{\cD_0}N$ for some
operator $ N$.

Recall that we say the long operator $ L:\ce^a\to \ce_a[-4]$ has
harmless  kernel if $ \ker(L)\subset \ker(\cD_0)$.
\begin{theorem}
  Suppose $ M$ is a conformally flat compact oriented manifold and
  that $ L$ has  harmless kernel.  Then $\ker(F:\ce_{(ab)_0}[2]\to
  \cF^{\bar{A}}{}_{b}[-3])$ is a conformally invariant subspace of
  $\ce_{(ab)_0}[2] $ isomorphic to the first cohomology of the
  deformation complex.
\end{theorem}
\noindent {\bf Proof:} The argument is formally almost identical to the
proof of theorem \ref{confhodge}. Suppose $ h$ is in the null space of
$ F$.  Then clearly $\BI F=B$ annihilates $ h$. But since this has the
form $ B=\bar{\cD_1}\cD_1$ it follows that $ (\cD_1 h, \cD_1 h)=0$. The
inner product is definite on $ \cW$ so we have $ h\in \ker(\cD_1)$.
So there is a map from $ \ker(F)$ to the first cohomology given simply by
$ h\mapsto [h]$.

In a parallel to the de Rham case, to invert this map we establish
that there is a unique element $h'\in \ker(F)$ in the class $[h]$. This
time we have $ h'=h+\cD_0 t $ for some tangent vector field $ t$. The
class is a subspace of $ \ker(B)$ so this boils down to solving $
{\gG} h+{\gG}\cD_0 t=0 $ or in other words
\begin{equation}\label{it}
 {\gG} h=-L t.
\end{equation}
Now since $ L$ is formally-self-adjoint and elliptic with harmless
kernel it follows that $ \image(L)$ is just the subspace in $
\ce_a[-4]$ orthogonal to the space of conformal Killing vectors in $
\ce^a$.  Since $ {\gG} h$ has the form $\bar{\cD_0} Nh$ it is
immediate that this lies in this image and so \nn{it} is solvable and
determines $ \cD_0 t$ uniquely as required.  \quad $ \Box$

In a choice of scale, $[F h ]_g $ has the form $ (B h, \gG)$. Thus the
conformally invariant space $\ker(F)$ is recovered in any choice of
scale by $ \ker(B) \cap \ker(\gG)$.

\section{Duality and final remarks}

Note that the $ H^3$ of the deformation complex is naturally the
vector space dual of $ H^1$. We have the conformally invariant map
$ H^3\to (H^1)^*$ given by $ U\mapsto (U,\cdot) $,
for $ U$ any
representative of $ H^3$.
That this is bijective
follows easily by choosing a metric from conformal class and using standard
Hodge theory arguments.
An analogous argument shows $ H^{4}=(H^0)^*$.
$H^0 $ is the cohomology at $ \ce^a$ of the complex $ 0\to \ce^a\to
\ce_{(ab)_0}[2]\to \cdots$. That is it is the vector space $\bW$
from above. Then $ H^4$ is invariantly realised as $\bW^*$ by
using the conformally invariant inner product to pair sections of $
\ce_a[-4]$ against conformal Killing vectors. From this final point we
note that the deformation resolution from the previous section could be
adjusted in a natural and conformally invariant way to the
``duality-symmetric''
sequence
$$
0\to\bW\to \ce^a\to \ce_{(ab)_0}[2]\to \cW \to \ce_{(ab)_0}[-2]\to
\ce_a[-4]\to\bW^*\to 0 .
$$
Of course similar remarks apply to the de Rham resolution.
By the invariant inner product of that case there is a conformally
invariant interpretation of $ H^{4-i}(M)$ as the vector space dual of
$ H^i(M)$.

Some further remarks: Firstly the operator $F$ in theorem
\ref{Bcurved} is not unique. There are many ways to modify the
formula. (For example, in an appropriate sense one can swap the order
of $\AD $ and $\BD $.)  Part of the motivation for the approach taken
was to make the operator $ \bB$ formally-self-adjoint by construction.
We have computed a ``conventional'' tensorial expression for the operator
of the theorem which differs from a constant multiple of the
linearised Bach operator by a (nontrivial)
conformally invariant lower order operator.
When one leaves the friendly confines of the de Rham complex, there
is much more ``room'' for different tractor constructions of invariant
operators to differ below the leading order.
At the date of writing, the precise meaning of the difference operator
described above is not clear.

On structures with vanishing Bach tensor the linearised Bach operator
annihilates the image of the Killing operator. It may be that this is
a good setting to generalise the ideas of Section \ref{defcx}.

We have already mentioned that much of the story told here extends to
other dimensions and \cite{fup,BrGometric} describe this. In fact
there is a large class of so called Bernstein-Gelfand-Gelfand sequences
(as in e.g.\ \cite{calderTD} and references therein) for which many of
these ideas extend. We mean here in the conformal setting but also to
some extent for other similar (i.e. parabolic) geometries.

\end{document}